\documentclass[11pt,a4paper]{article}
\usepackage[hyperref]{emnlp2020}
\usepackage{times}
\usepackage{latexsym}

\usepackage{multirow}
\usepackage{multicol}
\usepackage{booktabs}
\usepackage{graphicx}
\usepackage{amsmath}
\usepackage{float}

\usepackage{microtype}

 \aclfinalcopy % Uncomment this line for the final submission
\title{AWS CORD-19 Search: A Neural Search Engine for COVID-19 Literature}

\date{}

 \author{Parminder Bhatia, Lan Liu, Kristjan Arumae, Nima Pourdamghani, Suyog Deshpande,  \\ \textbf{Ben Snively, Mona Mona, Colby Wise,} 
 \textbf{George Price, Shyam Ramaswamy, Xiaofei Ma,}\\
 \textbf{Ramesh Nallapati, Zhiheng Huang, Bing Xiang, Taha Kass-Hout}\\
   Amazon Web Services AI\\
   \texttt{parmib, liuall, arumae, nimpourd, suyogd, snivelyb, }\\
   \texttt{monamo, colbywi, gwprice, shyar, xiaofeim, rnallapa, }\\ 
   \texttt{zhiheng, bxiang, tahak @amazon.com}}
   
\begin{document}
\maketitle
\begin{abstract}
Coronavirus disease (COVID-19) has been declared as a pandemic by WHO with thousands of cases being reported each day.
Numerous scientific articles are being published on the disease raising the need for a service which can organize, and query them in a reliable fashion. 
To support this cause we present AWS CORD-19 Search (ACS), a public, COVID-19 specific, neural search engine that is powered by several machine learning systems to support natural language based searches.
ACS with capabilities such as document ranking, passage ranking, question answering and topic classification  provides a scalable solution to COVID-19 researchers and policy makers in their search and discovery for answers to high priority scientific questions.
We present a quantitative evaluation and qualitative analysis of the system against other leading COVID-19 search platforms.
ACS is top performing across these systems yielding quality results which we detail with relevant examples in this work.

\end{abstract}

\section{Introduction}
\label{intro}

With the global outbreak of Coronavirus disease (COVID-19) \cite{doi:10.1056/NEJMoa2002032}, the world is in turmoil.  
Medical researchers are required to work quickly to fully understand and to provide a form of intervention for the virus.
Due to a large research focus on the disease, knowledge is published at a rapid rate throughout the world.  
One such repository of information is curated through the COVID-19 Open Research Dataset Challenge (CORD-19) \cite{wang2020cord}.
CORD-19 is a joint challenge put forth by Allen Institute (AI2), National Institutes of Health (NIH), and the United States federal government via the White House.  
The objective of the challenge is to make sense of and extract useful knowledge across thousands of scholarly articles related to COVID-19.
CORD-19 aims to connect the machine learning community with biomedical domain experts and policy makers in a race to identify effective treatments and management policies for COVID-19.
In accordance with this initiative our goal is to provide a scalable solution aimed at aiding COVID-19 researchers and policy makers in their search and discovery for answers to high priority scientific questions.
These questions should be understood in their natural language form; examples include: ``What do we know about COVID-19 risk factors?'', ``Which medications were most beneficial in the 2002 SARS outbreak?'', and ``Which is the most referenced paper doing study for Hydroxy-chloroquine?''
To appropriately answer these questions we require a system with a strong biomedical understanding of the natural queries and structured knowledge \cite{rotmensch2017learning}.

% \begin{figure}[t]
% \centering
% \resizebox{\columnwidth}{!}{
% \scriptsize
% % \multicolumn{2}{c}{\includegraphics[scale=0.55]{acl2020-templates/images/mainlogo.png}}\\
% \begin{tabular}{p{0.2\linewidth}p{0.6\linewidth}}
% \toprule
% \textit{query} & ``Are IL-6 inhibitors key to COVID-19?''\\
% \midrule
% \textit{response} & SARS-CoV-2 and \textbf{COVID-19}: is interleukin-6 (\textbf{IL-6}) the 'culprit lesion' of ARDS onset?\\
% \textit{research article} & ``...another monoclonal antibody against IL-6,  is being tested in a clinical trial against \textbf{COVID-19} (Sarilimumab \textbf{COVID-19}). Another drug that  showed  potential inhibition of \textbf{IL-6} related  JAK/STAT pathway is glatiramer acetate which  showed potential to downregulate both IL-17 and \textbf{IL-6}..''\\
% \textit{topic(s)} & Clinical Treatment \\
% \bottomrule
% \end{tabular}}
% \caption{A natural language query and response using AWS CORD-19 Search.}
% \label{figure:motivation}
% \end{figure}

\begin{figure*}[t]
\centering
\resizebox{\textwidth}{!}{
\tiny
\begin{tabular}{p{0.15\textwidth}p{0.15\textwidth}p{0.3\textwidth}p{0.1\textwidth}}
\toprule
Query & Article & Response & Topic(s) \\
\midrule
``Are IL-6 inhibitors key to COVID-19?'' & SARS-CoV-2 and \textbf{COVID-19}: is interleukin-6 (\textbf{IL-6}) the 'culprit lesion' of ARDS onset? & ``...monoclonal antibody against IL-6,  is being tested in a clinical trial against \textbf{COVID-19} (Sarilimumab \textbf{COVID-19}). Another drug that  showed  potential inhibition of \textbf{IL-6} related  JAK/STAT pathway is glatiramer acetate which  showed potential to downregulate both IL-17 and \textbf{IL-6}'' & Clinical Treatment \\\\
``When is the salivary viral load highest for COVID-19?'' & Elective, Non-urgent Procedures and Aesthetic Surgery in the Wake of SARS–COVID-19 [...] & ``Patients with COVID-19 have demonstrated \textbf{high viral loads} in the upper respiratory tract soon after their infection, with the \textbf{highest load} assumed to be the day before symptoms appear.'' & Clinical Treatment\\\\
``Is convalescent plasma therapy a precursor to vaccine?'' & COVID-19 convalescent plasma transfusion & ``...a passive immunotherapy, has been used as a possible therapeutic option when no proven specific \textbf{vaccine} or drug is available for emerging infections. '' & Clinical Treatment, Immunology, Lab Trials\\
\bottomrule
\end{tabular}}
\caption{Sample natural language queries and responses using AWS CORD-19 Search.  The response field is taken directly from the top result of this service.  Also provided are the article titles where the answer is taken from as well as selected topics for the response.}
\label{figure:motivation}
\end{figure*}

% \begin{figure*}[t]
%     \centering
%     \includegraphics[scale=0.205]{demo2.png}
%     \centering
%     \caption{AWS CORD-19 Search results page.}
%     \label{fig:demo}
% \end{figure*}

AWS CORD-19 Search (ACS) provides an easy to use search interface where researchers query using natural language \footnote{\url{https://cord19.aws/}}.
% such as, “When is the salivary viral load highest for COVID-19?” and “Is convalescent plasma therapy a precursor to vaccine?” CORD-19 Search produces precise answers as well as source documents. 
ACS goes beyond keyword matching by understanding question semantics to efficiently find relevant answers.
As illustrated in Figure \ref{figure:motivation}, we provide the system with a natural language query inquiring about \textit{IL-6 inhibitors} and showcase the system response with relevant query components highlighted.
This query can confuse a traditional search engine that relies solely on text matching as it can only capture term overlap between a query and a document and may not necessarily be relevant to the researcher’s true intent (i.e. learning implicit relations).
Our system, however, establishes the relationship between IL-6 and severity of SARS-CoV-2 showing evidence where elevated IL-6 occur in a large number of patients with severe COVID-19 and higher mortality rates. 
% Further researcher might be interested in “What is the mechanism of action of IL-6?” 
% CORD-19 understands the query and presents the mechanism of action behind these inhibitors showing “SARS-CoV-2 N protein activates IL-6 gene expression by binding directly or indirectly to NF-κB regulatory element on IL-6 promoter…”

In the following example, if tasked with understanding the epidemiology and transmission of COVID-19 a researcher may ask about \textit{salivary viral load}. 
We provide a further advantage from an organizational perspective using topics with domain specific relations.
The user can select \textit{clinical-treatments} to refine search articles to understand that ``...the highest load [is] assumed to be the day before symptoms appear.''
Similarly if commissioned to understand \textit{convalescent plasma therapy} and selecting the three topics as shown in the last example the system identifies the most relevant article and answers the question in a highlight.
% \iffalse added by Lan \fi
There are a few search systems have been developed to support COVID-19 search leveraging CORD-19 corpus. 
We evaluate the ACS against two other search engines, and ACS is proved to be a top-performing system. 
% \iffalse added by Lan \fi

In the following sections we will discuss the various individual Amazon Web Services (AWS) products that allow ACS to provide this functionality, highlighting the value to scientists who can quickly query, validate their research, and advance their investigations.

\begin{figure*}[t]
    \centering
    \includegraphics[scale=0.5]{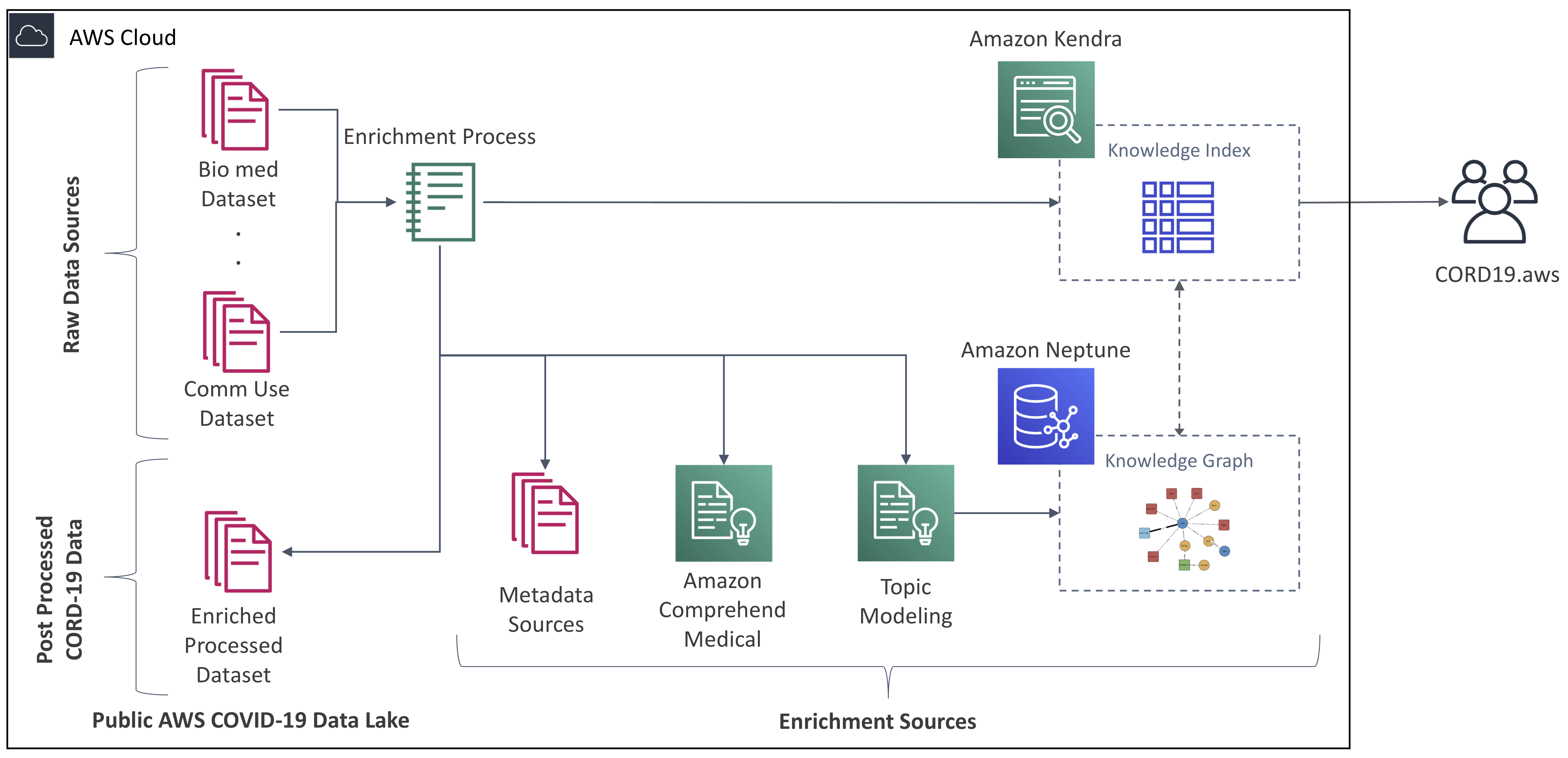}
    \centering
    \caption{System Architecture.}
    \label{fig:sys_arch}
\end{figure*}

\vspace{-0.07in}
\section{System Overview}
\vspace{-0.10in}
AWS CORD-19 Search (ACS) relies on a deep semantic search model to return a ranked list of relevant documents. 
This system performs document ranking, passage ranking, question answering and FAQ matching. 
Additionally, it leverages knowledge graphs and topic modeling to enrich the biomedical data and boost searching performance.  
% Amazon CORD-19 is based on a deep learning-based semantic search model to return a ranked list of relevant documents. 
% This system makes use of topic modeling, knowledge graphs, and natural language queries.
% Furthermore it leverages document ranking, reading comprehension, as well as FAQ matching.
This provides a scalable solution to COVID-19 researchers and policy makers.
In this section we present the overall architecture of the system and a closer look at several individual components.
Figure \ref{fig:sys_arch} provides an overview of this architecture.
\vspace{-0.1in}

\subsection{Amazon Kendra}
Amazon Kendra\footnote{\url{https://aws.amazon.com/kendra/}} is a semantic search and question answering service provided by AWS for enterprise customers. Kendra allows its customers to power  natural language based searching across their own data. 
As response to the worldwide COVID-19 pandemic, Kendra has also been tooled to support COVID-19 related searching and question answering using the document corpus from CORD-19. The end-to-end Kendra system consists of several components.

To complement the extracted answers, Kendra uses a deep learning based semantic search model to return a ranked list of relevant documents. 
Amazon Kendra's ability to understand natural language questions is at the core of its search engine returning the most relevant passage and related documents

\begin{figure*}[t]
    \centering
    \includegraphics[scale=0.65]{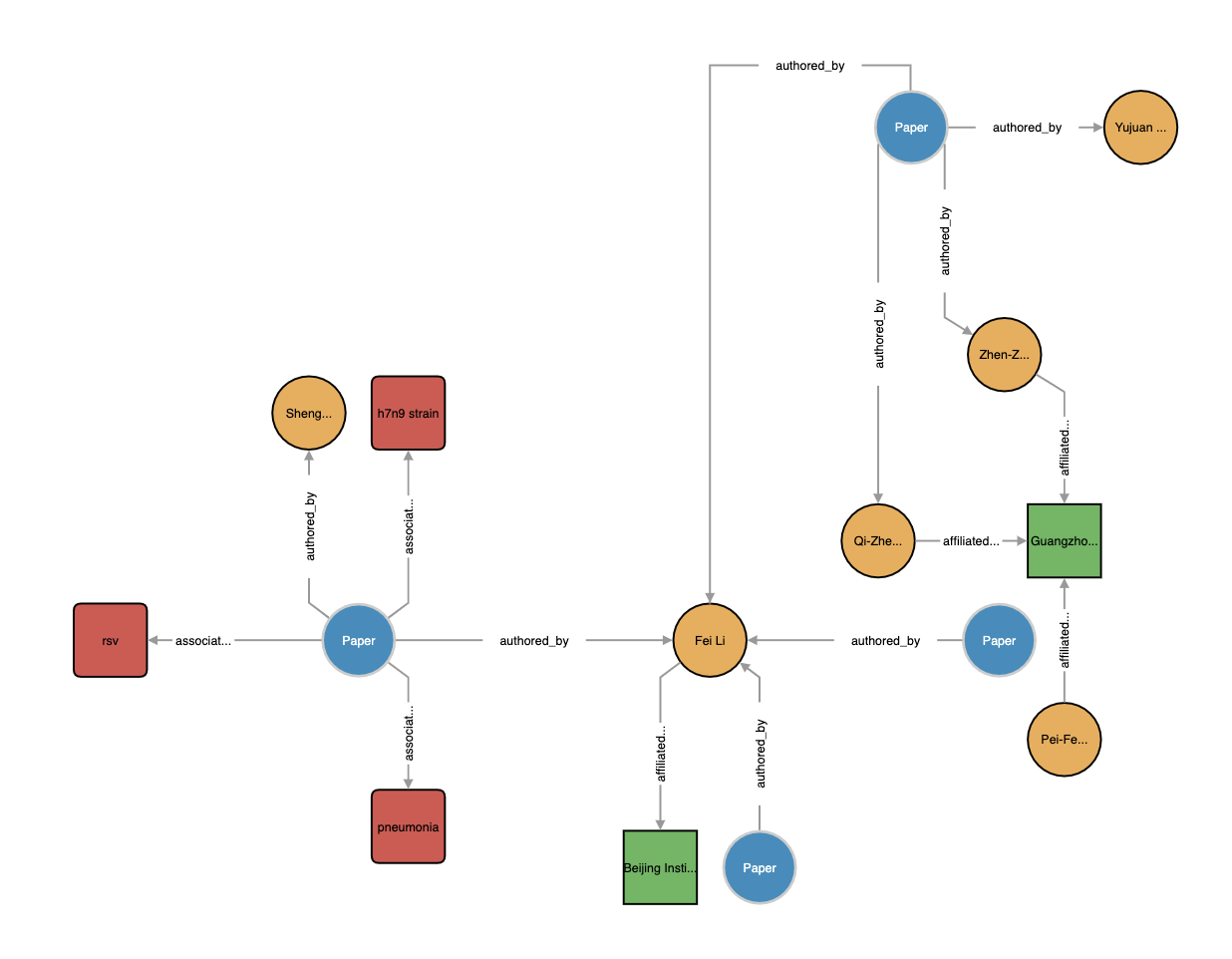}
    \centering
    \caption{Visualization of COVID-19 Knowledge graph.}
    \label{fig:ckg}
\end{figure*}

\begin{itemize}
%\small
    \item \textbf{Document ranking (DR)} -  Like any traditional search engine, Kendra returns a ranked list of relevant documents based on the user's query to fulfill their information needs. A deep semantic search model is used to understand natural language questions in addition to the keyword search.
    %The model follows a multi-stage search architecture that is composed of initial retrieval phase based on traditional term matching and a subsequent BERT-style model that reranks the candidates. This architecture facilitates Kendra to support search with both keyword queries and natural language questions.
    \item \textbf{Passage ranking (PR) \& Question Answering (QA)} - Kendra ranks the passages and tries to extract the answer from the top relevant passages with a deep reading comprehension model.
    \item \textbf{FAQ Matching (FAQM)} - If there exists Frequently Answered Questions and their corresponding answers, Kendra will automatically match a new coming query with FAQs and extract the answer if a strong match is found.
\end{itemize}

In order to improve Kendra CORD-19 search and make it clinically more relevant for medical researchers, we leverage knowledge extracted using the Amazon Comprehend Medical (CM) core NERe service as well as topics created from Amazon Comprehend Custom Classification  \footnote{\url{https://docs.aws.amazon.com/comprehend/latest/dg/how-document-classification.html}}. 
More explicitly, a knowledge graph is built using the medical entities extracted with CM NERe API, and the topics of the article are produced using a semi-supervised prior based LDA approach. 
Both are used to enrich the data when indexing Kendra.

\subsection{Comprehend Medical}
Amazon Comprehend Medical \footnote{\url{https://aws.amazon.com/comprehend/medical/}} (CM) \cite{8999113}, is a HIPAA eligible AWS service for medical domain entity recognition \cite{bhatia2018joint}, relationship extraction \cite{singh2019relation} and normalization. 
Comprehend Medical supports entity types divided into five different categories (Anatomy, Medical Condition, Medication, Protected Health Information, and Test, Treatment, \& Procedure) and four traits (Negation, Diagnosis, Sign and Symptom). 
These entities are directly used to enrich the Kendra search.

%%%%%%%%%%%%%%%%%%%%%%%%%%%%%%%%%%%%%%%%%%%%%%%%%%%%%%%%%%%%%%%%%%%%%%%%%%%%%%%%%%%%%%%%%%%%%%%%%%%%%%%%%
%%%%%%%%%%%%%%%%%%%%%%%%%%%%%%%%%%Added By Lan %%%%%%%%%%%%%%%%%%%%%%%%%%%%%%%%%%%%%%%%%%%%%%%%%%%%%%%%%%
% \begin{figure*}[t]
% \vspace{-0.12in}
% \centering
% \resizebox{\textwidth}{!}{
% \tiny
% \begin{tabular}{p{0.06\textwidth}p{0.45\textwidth}}
%     \toprule
%     \multicolumn{2}{l}{TREC-COVID}  \\
%     \midrule
%     query:          & serological tests for coronavirus \\
%     question:      & are there serological tests that detect antibodies to coronavirus? \\
%     narrative:  & Looking for assays that measure immune response to COVID-19 that will help determine
% past infection and subsequent possible immunity. \\
%     \midrule
%     \multicolumn{2}{l}{CovidQA} \\
%     \midrule
%     question:          & How does seasonality affect the transmission of COVID-19? \\
%     \bottomrule
%     \label{table:dataset}
% \end{tabular}}
% \vspace{-0.15in}
% \caption{Example of TREC-COVID and CovidQA}
% \label{figure:QA example}
% %\vspace{0.2in}
% \end{figure*}

\begin{table*}[t]
    \centering
    \begin{tabular}{cccccccc}
    \toprule
    Search Engine   & P@1 & P@5 & P@10 & P@20 & R@10 & R@20 & ndcg@20   \\
    \toprule
    \multicolumn{8}{c}{Keyword Queries} \\
    \midrule
    ACS   & 0.5250 & \textbf{0.5650} & \textbf{0.5325} & \textbf{0.4775} & \textbf{0.0260} & \textbf{0.0459} & \textbf{0.4380} \\
    Covidex   & 0.3421  & 0.2316 & 0.2079 & 0.1658 & 0.0109 & 0.0173 & 0.1633 \\
    COVID-19 RE   & \textbf{0.5750} & \textbf{0.5650} & 0.4775 & 0.4412 & 0.0236 & 0.0429 & 0.4022\\
    \toprule
    \multicolumn{8}{c}{Natural Language Questions} \\
    \midrule
    ACS   & \textbf{0.8750} & \textbf{0.7000} & \textbf{0.6400} & \textbf{0.5550} & \textbf{0.0345} & \textbf{0.0582} & \textbf{0.5357} \\
    Covidex   & 0.4750  & 0.4800 & 0.4225 & 0.3625 & 0.0204 & 0.0356 & 0.3229 \\
    COVID-19 RE   & 0.6000 & 0.5300 & 0.4925 & 0.4600 & 0.0267 & 0.0474 & 0.4133 \\
    % \toprule
    % \multicolumn{8}{l}{Overall} \\
    % \midrule    
    % ACS   & \textbf{0.7000} & \textbf{0.6325} & \textbf{0.5162} & \textbf{0.3282} & \textbf{0.0303} & \textbf{0.0790} & \textbf{0.4869} \\
    % Covidex   & 0.4103  & 0.3590 & 0.2667 & 0.1449 & 0.0158 & 0.0347 & 0.2451 \\
    % COVID-19 RE   & 0.5875 & 0.5475 & 0.4506 & 0.2770 & 0.0252 & 0.0677 & 0.4077 \\
    \bottomrule
    \end{tabular}
    \caption{Evaluation results on  TREC-COVID dataset} 
%    \vspace{-0.10in}
    \label{table:DR}
\end{table*}

%%%%%%%%%%%%%%%%%%%%%%%%%%%%%%%%%%Added By Lan %%%%%%%%%%%%%%%%%%%%%%%%%%%%%%%%%%%%%%%%%%%%%%%%%%%%%%%%%%
%%%%%%%%%%%%%%%%%%%%%%%%%%%%%%%%%%%%%%%%%%%%%%%%%%%%%%%%%%%%%%%%%%%%%%%%%%%%%%%%%%%%%%%%%%%%%%%%%%%%%%%%%

\subsection{COVID-19 Knowledge Graph}
Knowledge graphs (KGs) are structural representations of relations between real-world entities in the form of triplets containing a head entity, a tail entity, and the relation type connecting them. KG based information retrieval has shown great success in the past decades~\cite{dalton2014entity}. 
The COVID-19 Knowledge Graph (CKG) (Fig \ref{fig:ckg})  is a directed property graph constructed from the CORD-19 Open Research Dataset of scholarly articles. 
Entities including scholarly articles, authors, author institutions, citations, extracted topics and comprehend medical entities are used to form relations in the CKG. 
The resulting KG continues to grow as the CORD-19 dataset increases and currently contains over 335k entities and 3.3M relations. 
The CKG powers a number of features on ACS including: article recommendations, citation-based navigation, and search result ranking by author or institution publication count. 
Scientific article recommendations are made possible by a document similarity engine that quantifies similarity between documents by combining semantic embeddings obtained from a pre-trained language model~\cite{beltagy2019scibert} with document knowledge graph embeddings~\cite{wang2017knowledge, zheng2020dgl} capturing topological information from the CKG.

\subsection{Topic Models}
Topic modeling is a statistical discovery paradigm for generating topics that occur in a collection of documents.  
Perhaps the most widely used model for topic modeling is Latent Dirichlet Allocation (LDA) \cite{blei2003latent}, a generative model which groups documents together by observed content, often giving each document a mixture of topics it belongs to.  
An extension of this work termed Z-label LDA \cite{andrzejewski2009latent} utilizes priors to allow the model to force certain topics which the users have manually curated, or wish to see clustered together.  

\subsubsection{Generating Topics}
For the purposes of this work we experimented with 5, 10, and 20 topic models \footnote{These models were trained using CORD-19 data available as of April 6th, 2020.}.
The outputs of each clustering size were manually inspected and topic labels were provided by us when inspecting the top ten terms for each cluster.  
The final granularity of the topic models was chosen by manually deleting and merging topics from the 20 topic model.  
In general we were able to clearly extract groups which centered around important topics including virology, proteomics, epidemiology, and cellular biology to name a few.  
However when faced with 20 topics the less populated ones tended to be noisy, and captured peripheral information present in the input, such as language (e.g. Spanish and French) or provide redundancies with existing topics (e.g. two topics for Influenza).  
As a control we ran a publicly available implementation of Z-label LDA \footnote{\url{http://pages.cs.wisc.edu/~andrzeje/research/zl_lda.html}} with no priors which yields topics close to those extracted using Comprehend Topic Modeling.  
Although similar we observed better definition in certain groups (such as pulmonary diseases, and policy/industry), and decided to use this as the curation entry-point.  
Our goal was to limit these topics to ten, and compile them in advance as much as possible.  With the help of medical professionals we eliminated and combined topics to form the following: Vaccines/immunology, Genomics, Public health Policies, Epidemiology, Clinical Treatment, Virology, Influenza, Healthcare Industry, Pulmonary Infections, and Lab Trials (human).

% \begin{multicols}{2}
% \begin{small}
% \begin{enumerate}
%     \item  Vaccines/immunology
%     \item Genomics
%     \item Public health policies
%     \item Epidemiology
%     \item Clinical treatment
%     \item Virology
%     \item Influenza
%     \item Healthcare industry
%     \item Pulmonary infections
%     \item Lab trials (human)
% \end{enumerate}
% \end{small}
% \end{multicols}

%%%%%%%%%%%%%%%%%%%%%%%%%%%%%%%%%%%%%%%%%%%%%%%%%%%%%%%%%%%%%%%%%%%%%%%%%%%%%%%%%%%%%%%%%%%%%%%%%%%%%%%%%
%%%%%%%%%%%%%%%%%%%%%%%%%%%%%%%%%%Added By Lan %%%%%%%%%%%%%%%%%%%%%%%%%%%%%%%%%%%%%%%%%%%%%%%%%%%%%%%%%%

%\vspace{-0.08in}
%\vspace{-0.01in}
%%%%%%%%%%%%%%%%%%%%%%%%%%%%%%%%%%Added By Lan %%%%%%%%%%%%%%%%%%%%%%%%%%%%%%%%%%%%%%%%%%%%%%%%%%%%%%%%%%
%%%%%%%%%%%%%%%%%%%%%%%%%%%%%%%%%%%%%%%%%%%%%%%%%%%%%%%%%%%%%%%%%%%%%%%%%%%%%%%%%%%%%%%%%%%%%%%%%%%%%%%%%

\subsubsection{Multi-Label Classification}
Having to manually feed a topic model and re-train on the entire corpus once new data becomes available is largely inefficient. 
We therefore used the topic model labels to train a multi-label classifier \cite{read2011classifier}.
To evaluate the performance of this model we calculate the average F$_1$ across test samples by calculating the set overlap between our gold standard (topic model) and system labels (multi-label classification).
This held-out test set contains $20\%$ of the CORD-19 data available at the time.
%\begin{equation*}
   % \frac{1}{N}\sum_{n=1}^N F_1(y_n, \hat{y}_n),
    %\quad
    %F_1(y_n, \hat{y}_n) = 2 \cdot \frac{P \cdot R}{P + R + \epsilon}
    %\quad
    % P(y_n, \hat{y}_n) = \frac{| y_n \cap \hat{y}_n |}{\hat{y}_n}, 
    % \quad
    % R(y_n, \hat{y}_n) = \frac{| y_n \cap \hat{y}_n |}{\hat{y}_n}.
%\end{equation*}

\begin{table*}[t]
    \centering
    \begin{tabular}{l|cc|cc|cc}
    \toprule
        \multirow{2}{*}{Search Engine} & \multicolumn{2}{c|}{Top3} & \multicolumn{2}{c|}{Top10} & \multicolumn{2}{c}{Top30} \\ & EM & F1 &EM &F1 &EM &F1\\
    % \toprule
    %  & \multicolumn{2}{c}{Top3} & \multicolumn{2}{c}{Top30}\\
    % Search Engine   & EM & F1 & EM & F1 \\
    \midrule
    ACS   &\textbf{11.7} & \textbf{35.6} & \textbf{16.0} & \textbf{42.2} & \textbf{26.0} & \textbf{50.4} \\
    Covidex   & 0.90 & 18.9 & 2.10 & 24.2 & 3.60 & 27.6 \\
    COVID-19 RE  & 10.0 & 31.8 & 13.5 & 36.9 & 18.2 & 43.8 \\
    \bottomrule
    \end{tabular}
    \caption{Top results variation from KQ to NQ on TREC-COVID dataset} 
    %\vspace{-0.05in}
    \label{table:QueryVariation}
\end{table*}

\begin{table*}[t]
    \centering
    \begin{tabular}{l|ccc|ccc}
    \toprule
    
    \multirow{2}{*}{Search Engine} & \multicolumn{3}{c|}{PR} & \multicolumn{3}{c}{QA} \\ & P@1 & P@2 &P@3 &P@1 &P@2 &P@3 \\
    \midrule
    ACS          & \textbf{0.4074} & \textbf{0.5370} & \textbf{0.4938} & \textbf{0.3333} & \textbf{0.2593} & 0.2099 \\
    Covidex      & \textbf{0.4074}          & 0.4444 & 0.4074 & 0.1481 & 0.1852 & 0.2099 \\
    COVID-19 RE  & \textbf{0.4074}          & 0.3704 & 0.3580 & 0.2963 & 0.2407 & \textbf{0.2593}\\
    \bottomrule
    \end{tabular}
    \caption{Passage ranking (PR) and question answering (QA) performance on CovidQA dataset} 
    %\vspace{-0.05in}
    \label{table:PRQA}
\end{table*}

Using this metric our trained model achieved an average F$_1$ of $91.92$, with on average $2.37$ labels per document.  
Fewer than 1\% of the documents in the test set received no label, using $0.5$ as the confidence threshold.

\section{Evaluation}
%We observed the following key insights from our system.
% \begin{itemize}
% \vspace{-0.08in}
%     \item  ACS has the highest Top-3 Accuracy of question for each query
%     \vspace{-0.08in}
%     % \item The annotators have higher agreement using ACS than Covidex or the COVID-19 Research Explorer.
%     % \vspace{-0.08in}
%     %  \item  The F1 of selecting the top best answer for CORD-19 is 83.2 , comparing with 76.2 for Covidex and 69.8 for COVID-19 Research Explorer.
%     %   \vspace{-0.08in}
%     \item  The EM of selecting the top best answer for CORD-19 is 50.5, comparing with 41.5 for Covidex and 39.8 for COVID-19 Research Explorer.
% \end{itemize}

ACS supports document ranking (DR), passage ranking (PR) and question answering (QA) on the top suggested results. In this section we evaluate the overall performance of ACS with publicly available datasets. The TREC-COVID track \cite{voorhees2020arxiv} built DR test collections that contain document relevance ground truth and are used to assess DR only. For PR and QA, we manually annotate questions collected from CovidQA \cite{tang2020arxiv}, the first open sourced Covid-19 QA dataset. The ACS results are compared against Covidex\footnote{\url{https://covidex.ai/}} and Google Covid-19 Research Explorer\footnote{\url{https://covid19-research-explorer.appspot.com/}}  (COVID-19 RE), which are also COVID-19 public search engines that facilitate DR and answer highlighting in passage. 

The TREC-COVID challenge track contains 40 topic sets along with their document relevance judgement. The topic sets are written by its organizers with biomedical training, and motivated by search submitted to the National Library of Medicine and social media. Each topic consists of three fields with different levels of granularity, a keyword-based query (KQ), a more precise natural language question (NQ), and a longer descriptive narrative. The DR assessments are performed following the TREC pooling mechanism. The participants submitted ranked lists of documents for each topic set, based on which a depth of 10 to 20 documents are pooled and combined as a collection of $(q, D)$ pairs. 
The pairs are then assessed by annotators with in-domain expertise. 
There are three rounds of judgement available that corresponding to three different versions of CORD-19 corpus. To ensure sufficient coverage of annotation, we aggregate all rounds results into 33,064 $(q, D)$ relevance judgements. Since the document id may change across versions, we map ids from each round to the May 19 release of CORD-19 corpus. 

To collect DR results from the three systems, we crawled the top 50 articles by querying the engines with KQ and its NQ variation from the topic sets on June 15, 2020. The crawled data, which are characterized by the article title and link, are mapped to May 19 CORD-19 corpus. Note that articles that cannot be found in the corpus are removed to ensure fair comparison.

We use the standard DR metrics in our evaluation, namely, the precision and recall at $k$ (P@$k$, R@$k$) and normalized discounted cumulative gain in the top $k$ documents (NDCG@$k$). Note that we evaluate with $k$ up to 20 since TREC-COVID has a pooling depth with 20 at most.
Table \ref{table:DR} presents the DR performance of the engines over KQ and NQ, respectively. ACS performs consistently better than the other engines on NQ and mostly on KQ, and all engines perform better on NQ comparing with KQ.

In addition, we are able to evaluate how robust each system is against query variation from KQ to NQ. 
Ideally, the results shall remain unchanged with query variation that requests the same information. 
We define exact match (EM) and F1 score among top $k$ results to evaluate the robustness. Let $Q$ be the topic sets, and $NQ_q^k$, $KQ_q^k$ denote the top $k$ searching results of natural language question and keyword query corresponding to the same topic $q\in Q$, respectively. As shown in Eq.(\ref{equation:1}), we take the top $k$ results $KQ_q^k$ as ground truth, and compute the average exact match and F1 score of the top k results $NQ_q^k$, and then average over all queries. The article title string is used as comparison key, and EM and F1 are standard that the maximum is taken over all the ground truth articles. 
Table \ref{table:QueryVariation} demonstrates that ACS has the best capability to provide consistent results with query variation.

% By leveraging the article title string as the comparison key, and the top k results of KQ search as ground truth, we compute the exact match (EM) and F1 score of the top k NQ results averaging over all queries. 
\begin{figure*}[t]
%\vspace{-0.12in}
\centering
\resizebox{\textwidth}{!}{
\tiny
\begin{tabular}{p{0.2\textwidth}p{0.2\textwidth}p{0.2\textwidth}}
\toprule
\multicolumn{1}{c}{ACS} & \multicolumn{1}{c}{Covidex} & \multicolumn{1}{c}{COVID-19 RE} \\
\toprule
\multicolumn{3}{c}{What is the incubation period of the virus?} \\
\midrule
Importance of Social Distancing: Modeling the spread of 2019-nCoV using Susceptible-Infected-Quarantined-Recovered-t model & Deadly viral syndrome mimics & Prediction of the virus incubation period for COVID-19 and future outbreaks \\\\

``Studies on the nature of the \textbf{virus} have suggested different \textbf{incubation periods} of the \textbf{virus}, and reports have suggested a median \textbf{incubation period of 5-6 days} and a very high symptom probability \textbf{period} of 14 days [5] '' & ``...\textbf{The incubation period is typically 3 to 14 days with the symptoms of an acute nonspecific, flulike illness developing suddenly}...\textbf{The incubation period is 7 to 10 days before the onset of symptoms [37]. This incubation period provides the potential for worldwide exposure because a person harboring the virus can expose the world at large via air travel}...'' & ``…while minimizing the negative consequences of the quarantine. 70 \textbf{71 The length of the incubation period varies both across and within virus families 4 . To our knowledge, 72 genomic features (if any) that correlate with the incubation}…''\\
\bottomrule
\end{tabular}}
\caption{Top-1 result of article title and displayed passage by querying 'What is the incubation period of virus?'}
\label{figure:QA example}
\vspace{0.2in}
\end{figure*}
\begin{align}
\begin{split}
& EM(NQ, KQ, k) \\
= & \dfrac{1}{|Q|}\sum_{q\in Q}\dfrac{1}{k}\sum_{i=1}^k em(NQ_q^k(i), KQ_q^k) \\
\label{equation:1}
\end{split}
% \begin{split}
% & F1(KQ, NQ, k) \\
% = & \dfrac{1}{|Q|\cdot k}\sum_{q\in Q}\sum_{i=1}^k f1(NQ_q^k(i), KQ_q^k) \\
% \label{equation:2}
% \end{split}
\end{align}

Next, we evaluate the performance of PR and QA with CovidQA dataset. CovidQA contains 27 questions and their answers from 124 question-article pairs, which are selected by in-domain volunteers as the most promising COVID-19 literature. Since the annotation does not possess sufficient answer coverage over the entire corpus, we leverage our internal annotation resources to make PR and QA judgement. More explicitly, we crawled the top 3 results characterized by the article, displayed passage and the highlighted text snippet on June 15, 2020 from ACS, Covidex and COVID-19 RE, respectively. After that, the annotators assess PR and QA in terms of whether the displayed passage contains relevant information and whether the highlighted text snippet answers the given question. To avoid bias, the crawled results are combined and shuffled randomly, therefore, the annotators do not have access to the source engine and the rank position information during the judgement. 

Table \ref{table:PRQA} presents precision of top PR and QA results. We use P@$k$ instead of EM and F1 to evaluate QA since ACS highlights the answer while Covidex and COVID-19 RE highlight the sentence that contains the answer. 
Instead of extracting the answer, it is more reasonable to judge whether the highlighted text answers the question and compute precision. ACS achieves better accuracy on both PR and QA on most metrics. Note that ACS highlights answer snippet for at most three passages \iffalse thresholding by the model confidence score \fi, while Covidex and COVID-19 RE highlight all displayed passages. This explains why ACS P@3 underperforms COVID-19 RE. Explicitly with an example, Figure \ref{figure:QA example} displays the topmost results of querying ``What is the incubation period of virus" from the three systems. ACS highlights the answer in the passage. In contrast, Covidex presents an article published at 2004 with information of virus which is irrelevant with coronavirus, and COVID-19 RE does not answer the question at all. 

As evident from the above results, ACS is one of the top-performing systems that provides high quality informative results over CORD-19 search.

%%%%%%%%%%%%%%%%%%%%%%%%%%%%%%%%%%Added By Lan %%%%%%%%%%%%%%%%%%%%%%%%%%%%%%%%%%%%%%%%%%%%%%%%%%%%%%%%%%
%%%%%%%%%%%%%%%%%%%%%%%%%%%%%%%%%%%%%%%%%%%%%%%%%%%%%%%%%%%%%%%%%%%%%%%%%%%%%%%%%%%%%%%%%%%%%%%%%%%%%%%%%

\begin{figure*}[t]
\centering
\resizebox{\textwidth}{!}{
\tiny
\begin{tabular}{p{0.2\textwidth}p{0.2\textwidth}p{0.3\textwidth}}
\toprule
Query & Article & Response \\
\midrule
``What \textit{medications} were most beneficial in the SARS outbreak?'' & Development of chemical inhibitors of the SARS coronavirus: Viral helicase as a potential target & ``...spread of SARS, a number of broad-spectrum antiviral medications were empirically administered to the SARS patients during the SARS outbreak in 2003. These medications include \textbf{ribavirin, HIV protease inhibitors, corticosteroids, and alpha-interferon (IFN-a)}. In a retrospective review of treatment...''\\\\
``What \textit{measures} were most beneficial in the SARS outbreak?'' & Impact of quarantine on the 2003 SARS outbreak: A retrospective modeling study & ``During the 2003 Severe Acute Respiratory Syndrome (SARS) outbreak, \textbf{traditional intervention measures such as quarantine and border control were found to be useful in containing the outbreak}...'' \\\\
``What did we learn from the SARS outbreak?'' (no topic) & Use of quarantine in the control of SARS in Singapore & ``The main lesson to learn from the SARS outbreak is the capability of an emerging infection to cause a pandemic in a short span of time and the paradigm shift needed to respond to such a disease.''\\\\
``What did we learn from the SARS outbreak?'' (clinical-treatment) & Characteristics of COVID-19 infection in Beijing & ``We compared the epidemic features between COVID-19 and 2003 SARS for learn lessons and control the outbreak.''\\
\bottomrule
\end{tabular}}
\caption{Sample queries demonstrating the semantic understanding of ACS, the use-fullness of Comprehend Medical NERe, and the utility of topic modeling for filtering.}
\label{figure:examples}
\end{figure*}

\section{Analysis}
In this section we look into a number of sample queries to shed light on how different components of ACS help in improving search results. 
We begin by observing how small semantic differences in the query alter the results.
The first sample in Figure \ref{figure:examples} is specific to medications.
While the top result does not include the term medication the system highlights \textit{ribavirin} and \textit{corticosteroids}.
The CORD-19 system understands that these terms represent medications with the help of CM NERe engine.
In the second example we change \textit{medications} to \textit{measures} and observe the top result discussing border control, and quarantine.
This clearly demonstrates that Amazon Kendra has a deep comprehension of token and query meanings. 

Finally, we take a look at the effects of topic modeling when grouping and filtering results.
The last two examples in Figure \ref{figure:examples} showcase the difference this makes in the top result.
Without specifying any topic the resulting article discusses high level policy, specifically quarantine measures in Singapore.
When we filter by clinical treatment the top result instead focuses on infections which is covered in the clinical setting.
Furthermore the extracted text returned to the user still focuses on lessons learned staying true to the query.

\section{Limitations and Future Directions}
\vspace{-0.02in}
AWS CORD-19 is an initial step towards helping medical researchers find relevant content in a timely and meaningful way. In order to improve the robustness, we see following areas as direction for future research.

\textbf{Feedback Loop} - Since ACS is a search engine the motivation would be to evaluate it as such; using well-established methodologies based on test collections—comprising topics (information needs) and human annotations. 
Since no designated evaluation data exist, our initial focus is to capture different interactions and feedback. 
Currently, ACS lacks the feedback loop and federated learning approaches where the system would continuously learn and improve the search. 
However, the system captures feedback from the researchers in the form of implicit and explicit reactions. 
Implicit feedback evaluation consists of topics of interests, their clicks as well as the ranking of the results which were selected by medical researchers. 
Explicit feedback evaluation is captured by providing up-down rating associated with each search results. 
In the future results can be personalized based on this feedback.
Now that we have a system in place, our efforts have shifted to broader engagement with potential stakeholders to solicit additional guidance, while trying to balance between the features and ranking.

\textbf{Q\&A Curation} - Curation and normalization of questions have potential a use-case of presenting trending questions asked by the medical research community at a particular point. 
However, curation would involve capturing the questions asked as well as identifying similar questions that can be later normalized. Currently, there is no mechanism to curate the questions asked by the researchers.

\textbf{Summarization} - Currently, ACS outputs the relevant passage based on the query. 
It would be beneficial to get the overall summary of the paper.
A potential future direction would be to generates summaries \cite{raffel2019exploring} from paper abstracts and full body.

\section{Conclusion}
\vspace{-0.02in}
This paper describes our efforts in building AWS CORD-19 Search with its capabilities consisting of topic based, knowledge graph, and natural language search queries.
This is further enhanced with reading comprehension and FAQ matching as well as document ranking providing a scalable solution to COVID-19 researchers and policy makers in their search and discovery for answers to high priority scientific questions. 
Our solution is powered by Amazon Kendra, Comprehend Medical and Neptune which incorporate the latest neural architectures to provide information access capabilities to the CORD-19 challenge. \iffalse added by Lan \fi By comparing with other public search engines that support similar functionalities over COVID-19 search, ACS is demonstrated to be powerful on its document ranking and question answering components. \iffalse added by Lan \fi We hope that our solution can prove useful in the fight against this global pandemic, and that the capabilities we have developed can be applied to analyzing the scientific literature more broadly.

\section*{Acknowledgments}
\vspace{-0.02in}
We acknowledge the broader collaboration with AI2 team and White house as well as the broader AWS CORD-19 team including Kendra Science team, Tyler Stepke, Kingston Bosco, Victor Wang, Vaibhav Chaddha, Miguel Calvo, Ninad Kulkani, Kevin Longofer, Ray Chang, Adrian Bordone, Tony Nguyen, and Kyle Johnson.

\bibliography{main}
\bibliographystyle{acl_natbib}

\end{document}